\documentclass[aps,prl,lengthcheck,superscriptaddress,amsmath,amssymb]{revtex4-2}
\usepackage{graphicx}
\usepackage{bm}
\usepackage{braket}
\usepackage{microtype}
\usepackage[colorlinks,allcolors=blue]{hyperref}
\usepackage{xcolor}

\def\equationautorefname#1#2\null{Eq.#1(#2\null)}

\renewcommand{\Re}{\operatorname{Re}}
\renewcommand{\Im}{\operatorname{Im}}

\begin{document}

\title{Electrostatic nature of cavity-mediated interactions between low-energy matter excitations}

\author{Petros-Andreas Pantazopoulos}
\email{petros.pantazopoulos@uam.es}
\affiliation{ Departamento de Física Teórica de la Materia Condensada and Condensed Matter Physics Center (IFIMAC), Universidad Autónoma de Madrid, E-28049 Madrid, Spain
}

\author{Johannes Feist}
\email{johannes.feist@uam.es}
\affiliation{ Departamento de Física Teórica de la Materia Condensada and Condensed Matter Physics Center (IFIMAC), Universidad Autónoma de Madrid, E-28049 Madrid, Spain
}
\author{Akashdeep Kamra}
\email{akashdeep.kamra@uam.es}
\affiliation{ Departamento de Física Teórica de la Materia Condensada and Condensed Matter Physics Center (IFIMAC), Universidad Autónoma de Madrid, E-28049 Madrid, Spain
}

\author{Francisco J. García-Vidal}
\email{fj.garcia@uam.es}
\affiliation{ Departamento de Física Teórica de la Materia Condensada and Condensed Matter Physics Center (IFIMAC), Universidad Autónoma de Madrid, E-28049 Madrid, Spain
}
\affiliation{ Institute of High Performance Computing, Agency for Science, Technology, and Research (A*STAR), Connexis, 138632 Singapore, Singapore}

\date{\today}

\begin{abstract}
The use of cavity quantum electrodynamical effects, i.e., of vacuum electromagnetic fields, to modify material properties in cavities has rapidly gained popularity and interest in the last few years. However, there is still a scarcity of general results that provide guidelines for intuitive understanding and limitations of what kind of effects can be achieved. We provide such a result for the effective interactions between low-energy matter excitations induced either directly by their mutual coupling to the cavity electromagnetic (EM) field or indirectly through coupling to mediator modes that couple to the EM field. We demonstrate that the induced interactions are purely electrostatic in nature and are thus fully described by the EM Green's function evaluated at zero frequency. Our findings imply that reduced models with one or a few cavity modes can easily give misleading results.
\end{abstract}

\maketitle

Over the last years, there has been increasing interest in the manipulation of matter with cavities ``in the dark'', i.e., in the absence of external driving, through the creation of hybrid quantum light-matter polaritonic states~\cite{garcia-vidal_manipulating_2021}, 
with experimental observations and theoretical predictions forming the emerging field of cavity quantum electrodynamic (QED) materials~\cite{schlawin_cavity_2022}. 
For resonant interactions, where matter excitations are energetically close to a cavity mode, the strong interaction with vacuum electromagnetic (EM) fields leads to the formation of polaritons and a plethora of effects takes place~\cite{garcia-vidal_manipulating_2021,schlawin_cavity_2022}.
For instance, the conductivity in molecular semiconductors can be increased by one order of magnitude~\cite{orgiu_conductivity_2015}, energy transport can be enhanced in both organic~\cite{feist_extraordinary_2015,schachenmayer_cavity-enhanced_2015} and inorganic materials~\cite{paravicini-bagliani_magneto-transport_2019} using excitonic states, which can also be used in the formation of superradiant excitonic insulators~\cite{mazza_superradiant_2019}, or for many-body synchronization dynamics~\cite{Yokoshi2017} or interaction enhancement by nonlocality and band degeneracy~\cite{Kinoshita2019}.
The breakdown of the topological protection in the integer quantum Hall effect has also been demonstrated by utilizing a 2D electron gas~\cite{appugliese_breakdown_2022}. Meanwhile, low-energy matter excitations have been also chosen to interact \emph{directly} but off-resonantly with confined EM modes for modifying superconductivity~\cite{schlawin_cavity-mediated_2019,curtis_cavity_2019}, ferromagnetism~\cite{roman-roche_photon_2021}, and ferroelectricity~\cite{ashida_quantum_2020,lenk_dynamical_2022}.

Alternatively, EM modes can be considered to couple with other bosonic modes (e.g., phonons) which in turn interact with the low-energy matter excitations. In this scheme, a direct coupling between the EM modes and the matter is presumably weak and thus disregarded, resulting in the EM modes mediating matter-matter interaction \emph{indirectly}.
Along this line, it has been proposed that cooperativity of optically active phonons and photons can alter the electron-phonon coupling and thus influence superconductivity~\cite{sentef_cavity_2018}.
In addition, it has been reported that the critical temperature of a conventional superconductor may be increased~\cite{hagenmuller_enhancement_2019} by 
strong coupling to surface plasmons and a long-range spin alignment can be favoured yielding enhanced magnetization~\cite{thomas_large_2021}.
At last, by employing a different scheme, it has been reported that confined photons coupled to spin excitations, the magnons, can also affect high-temperature superconductivity~\cite{curtis_cavity_2022}. 

Within this scheme of EM modes indirectly mediating matter-matter interactions, an intriguing scenario arises when the spatially extended EM modes resonantly hybridize with localized bosonic modes that couple with the similarly localized matter excitations. One may expect the hybrid polaritons thus formed to be more effective in mediating matter-matter interaction due to the polaritons' spatially extended character, inherited from the EM modes, and their direct coupling to matter, inherited from the localized bosons. Thus, one may expect an enhanced matter-matter interaction when the EM mode is resonant with the localized boson coupled directly with matter~\cite{thomas_large_2021}. At the same time, since the low-energy matter excitations are not resonant with any of the bosonic (including EM) modes in the system, one must consider all of the supported bosonic modes in evaluating the effective matter-matter interaction. Often, with the aim of capturing the key qualitative physics, previous theoretical works have considered one bosonic mode mediating the matter-matter interactions.

\begin{figure}
\centering
\includegraphics[width=\linewidth]{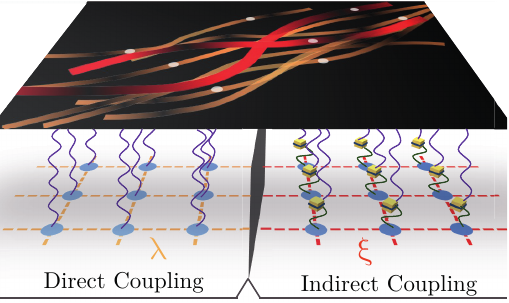}
\caption{Schematic illustration of vacuum EM modes (upper layer) interacting with low-energy matter excitations (bottom layer, denoted with circles), either directly (left panel) or indirectly (right panel) through a resonant local optically-active bosonic mediator (denoted with cubes).
The strings indicate the coupling between light, matter and the mediator.
Effective matter-matter interactions are induced by vacuum EM fields, which are quantified as $\lambda$ and $\xi$ for the two cases.}
\label{fig1}
\end{figure}

In this Letter, by studying the direct and indirect coupling scenarios for off-resonant interactions described above, we theoretically examine  how the mutual interaction of low-energy matter excitations is modified in the presence of quantized EM fields. 
First we analyze the situation in which these matter states are directly coupled to cavity EM modes, as schematically depicted in the left panel of \autoref{fig1}, leading to an effective interaction $\lambda$ between the matter components. 
The second scenario considers the presence of a localized mediator that is coupled to both the low-energy matter states of interest locally and to the surrounding EM field, as illustrated in the right panel of \autoref{fig1}, also giving rise to an effective interaction $\xi$. 
We find that in both cases, a consideration of all the EM modes supported by the cavity is mandatory to obtain reliable results.
By doing so, we demonstrate that the effective interactions, both $\lambda$ and $\xi$, have an electrostatic origin, i.e., they only depend on the zero-frequency response of the EM field in the cavity. We thus find that common models restricted to one or a few modes can produce misleading results. We also provide a simple physical interpretation of cavity-mediated effects based on electrostatic interactions. Furthermore, in contrast with the expectation described above, for the indirect coupling scenario outlined above, a resonance between the EM and localized bosonic modes is found not to enhance effective matter-matter interactions due to a nontrivial cancellation of contributions from the bonding and antibonding hybridized modes.

\emph{Direct matter-photon interaction}.---
We first consider the case in which a set of quantized EM modes interacts with low-energy matter excitations through the dipole--electric-field interaction in the long wavelength limit.
The corresponding Hamiltonian can be written as
\begin{equation}\label{eq:Htot}
H=H_{le}+\sum_{n}\hbar\omega_n a_n^\dagger a_n +\sum_{i,n}\widehat{\boldsymbol{\mu}}_i\cdot[\mathbf{E}_n(\mathbf{r}_i)a_n+\mathbf{E}_n^\ast(\mathbf{r}_i)a_n^\dagger].
\end{equation}
where $\widehat{\boldsymbol{\mu}}_i$ is the dipole operator of the $i$-th matter excitation, which is assumed to be spatially located at $\mathbf{r}_i$, and $a_n~(a_n^\dagger)$ is the annihilation (creation) operator associated with the EM mode of frequency $\omega_n$ and quantized electric field $\mathbf{E}_n(\mathbf{r})$. Notice that the light-matter interaction is treated beyond  the usual rotating wave approximation as we are dealing with off-resonant interactions.
The first term of the Hamiltonian, $H_{le}$, describes the matter excitations and/or direct matter-matter interactions, whose characteristic frequencies are assumed to be substantially lower than any significantly coupled cavity modes. 
The dipole self-energy term, which is one of the contributions to the free-state Lamb shift, is also included in $H_{le}$.

The effect of the EM environment on the matter states can be analyzed by deriving an effective Hamiltonian in which the photonic degrees of freedom are traced out~\cite{wang_phase_1973,hepp_superradiant_1973,coleman_introduction_2015,roman-roche_effective_2022}.
In the thermodynamic limit of many matter excitations, the desired effective Hamiltonian is obtained as
\begin{equation}\label{eq:Heff}
H_\mathrm{eff} =  H_{le}-\sum_{i,j} \widehat{\boldsymbol{\mu}}_i\cdot \boldsymbol{\lambda}_{ij} \cdot \widehat{\boldsymbol{\mu}}_j \;,
\end{equation}
where the effective coupling between matter states $i$ and $j$, $\boldsymbol{\lambda}_{ij}$, induced by the vacuum EM field is given by
\begin{equation}\label{eq:polaritonic_coupling}
\boldsymbol{\lambda}_{ij} = \sum_{n} \boldsymbol{\lambda}^{(n)}_{ij}= \sum_{n}\Re\left[\dfrac{\mathbf{E}_n(\mathbf{r}_i)\otimes\mathbf{E}^\ast_n(\mathbf{r}_j)}{\hbar\omega_n}\right] \;,
\end{equation} 
with $\otimes$ denoting the dyadic product of two vectors. A detailed derivation of this effective interaction can be found in the Supplemental Material \cite{supplemental}.

This result shows that an effective interaction between matter excitations $i$ and $j$ can indeed be induced by their common coupling to vacuum EM fields. 
Notice that this coupling is independent of the energy of the matter states, which is consistent with our initial assumption that the energies of the matter excitations are negligible compared to those of the EM modes. 
The tracing-out procedure of the EM degrees of freedom can be seen as an adiabatic elimination of the fast photon modes and, therefore, the energies of the matter excitations do not play any role in determining the strength of their effective interaction mediated by the vacuum EM field. 
As the effective coupling $\lambda_{ij}$ does not have a resonant nature, this already indicates that truncation of the sum over all the EM modes in \autoref{eq:polaritonic_coupling} to just a single mode (or several ones) could lead to incorrect results.

Furthermore, as we show in the following, the sum over modes can be performed explicitly by employing the macroscopic QED formalism~\cite{buhmann_dispersion_2012}, which provides a recipe for quantizing the EM field in any geometry, including with lossy materials. 
The quantized EM mode structure is fully encoded in the (classical) EM dyadic Green's function $\mathbf{G}(\mathbf{r}, \mathbf{r}',\omega)$, with the electric field operator becoming 
\begin{equation}
\mathbf{E}(\mathbf{r})= \sum_{p}\int_0^\infty \mathrm{d}\omega\int \mathrm{d}\mathbf{r}^\prime\mathbf{G}_p(\mathbf{r}, \mathbf{r}^\prime, \omega){\mathbf{f}}_p( \mathbf{r}^\prime, \omega) + \mathrm{H.c.}\label{eq:efield_qed}\;,
\end{equation}
where $\mathbf{f}_p(\mathbf{r}', \omega)$ are the bosonic operators of the medium-assisted EM modes, $p$ is an index labeling the electric and magnetic contributions, and $\mathbf{G}_p(\mathbf{r}, \mathbf{r}', \omega)$ are functions related to the dyadic Green's function. 
As detailed in the Supplemental Material, inserting this expansion in \autoref{eq:polaritonic_coupling} leads after some algebra (taking into account that the abstract sum over modes $n$ becomes a combination of sums and integrals) to
\begin{align}
\boldsymbol{\lambda}_{ij} &= \dfrac{1}{\pi\epsilon_0c_0^2}\int_0^\infty \mathrm{d}\omega \omega \Im\mathbf{G}(\mathbf{r}_i,\mathbf{r}_j,\omega)\;,
\end{align}
where we have used the Green's function identity $\sum_{p}\int \mathrm{d}\mathbf{s} \mathbf{G}_p(\mathbf{r}_i,\mathbf{s},\omega)\mathbf{G}_p^{*\mathrm{T}}(\mathbf{r}_j,\mathbf{s},\omega) = \frac{\hbar\omega^2}{\pi\epsilon_0c_0^2} \Im\mathbf{G}(\mathbf{r}_i,\mathbf{r}_j,\omega)$. Here, $\epsilon_0$ is the vacuum permittivity and $c_0$ is the speed of light in vacuum. 
Although the breakdown of the dipole approximation would in principle introduce a cutoff at high frequencies in Eq. (5), the integral converges very rapidly, as shown in the Supplemental Material, allowing to safely extend the upper frequency limit to infinity.
Using $\Im z = (z-z^\ast)/2i$ and the causality principle $\mathbf{G}(\mathbf{r}_i,\mathbf{r}_j,\omega)^\ast = \mathbf{G}(\mathbf{r}_i,\mathbf{r}_j,-\omega)$, this can be rewritten as an integral over the whole real line. 
Since $\omega {\mathbf{G}}(\mathbf{r}_i,\mathbf{r}_j,\omega)$ has a simple pole at $\omega=0$ and no other poles on the real axis or upper complex half space~\cite{novotny_principles_2012,buhmann_dispersion_2012}, contour integration yields the residue at $\omega=0$, i.e.,
\begin{equation}\label{eq:lambda_gf}
\boldsymbol{\lambda}_{ij} =\dfrac{1} {2\epsilon_0c_0^2}\left[\omega^2\mathbf{G}(\mathbf{r}_i,\mathbf{r}_j,\omega)\right]_{\omega=0}\;.
\end{equation}
This expression is a key finding of this Letter as it reveals that the effective interaction between low-energy matter excitations mediated by their direct coupling to the cavity EM modes is determined by the former's mutual electrostatic interaction. Note that this result applies for any material system, accounting also for a free-space EM environment. In this case, the corresponding Green's function, $\mathbf{G}_0$, feeding Eqs. (5) and (6) leads to the free-space electrostatic dipole-dipole interaction for $\boldsymbol{\lambda}_{ij}$, as described in the Supplemental Material. When dealing with a more complex EM environment such as a ``cavity'', which is characterized by a Green's function that incorporates scattering by the material,  $\mathbf{G}=\mathbf{G}_0+\mathbf{G}_S$, Eqs. (5) and (6) provide a simple recipe to evaluate cavity-modified matter interactions, and also imply that results obtained with theoretical treatments that include only one or a few EM modes when describing the vacuum EM field can potentially give misleading results.

The necessity of including the continuum of EM modes to evaluate off-resonant effects has long been known in the context of Casimir-Polder interactions~\cite{Casimir1948,buhmann_dispersion_2012}, and has been pointed out in the context of cavity-modified materials when treating the superradiant phase transition~\cite{de_bernardis_cavity_2018}, which can be understood as a ferroelectric instability with electrostatic characteristics, and when evaluating the possibility to achieve non-perturbative vacuum shifts~\cite{Saez-Blazquez2023}.

\textit{Indirect matter-photon interaction}.---
As commented above, an alternative mechanism to generate effective matter-matter interactions is to utilize a mediator mode which is strongly coupled to vacuum EM fields.
The key idea is that an optically active bosonic mode could couple to both the low-energy matter excitations and the EM modes, thus mediating their interaction. Furthermore, it can be speculated that if the energy of this bosonic mode resonates with that of one of the EM modes supported by the cavity, their hybridization could lead to a large long-range effective interaction between the matter states. 
To capture such a scenario, we treat a Hamiltonian
\begin{equation}\label{eq:Htot_ind}
H=H_{le}+H_p+H_m+H_{m-p}+H_{le-m}\;,
\end{equation}
which corresponds the addition of bosonic mediator modes $c_i$ of frequency $\Omega$ to \autoref{eq:Htot}. These modes are described by their bare Hamiltonian, $H_m=\sum_{i}\hbar\Omega c_i^\dagger c_i$, and their interaction with both cavity EM modes, $H_{m-p}$, and low-energy matter excitations, $H_{le-m}$.

Without loss of generality, we consider that the mediator bosonic modes are localized and couple to the cavity modes via the dipole--electric-field interaction
\begin{align}\label{eq:mediator-em-interaction}
H_{m-p}&=\sum_{i,n}\left(\zeta_{in}a_n c_i^\dagger+\zeta_{in}^\ast a_n^\dagger c_i\right)\;,
\end{align}
where $\zeta_{in}=\boldsymbol{\nu}_i^\ast\cdot\mathbf{E}_n(\mathbf{r}_i)$, $\boldsymbol{\nu}_i$ being the dipole moment associated with mediator mode $i$, which is assumed to be spatially located at $\mathbf{r}_i$. The on-resonant mediator-field interaction is here expressed within the rotating-wave-approximation.
Regarding $H_{le-m}$, we also assume that the coupling between matter states and the mediator is linear in the mediator operators $c_i$. 
Such a scheme can be realized, for example, for localized spins or electrons interacting with local dipole-active phonon modes that in turn couple to the EM field.
The mediator states alone are assumed to be local and thus do not induce long-range interaction between the low-energy excitations, and $H_{le-m}$ accounts for only local and independent coupling, $\Gamma_i$, which is  
\begin{align}\label{eq:Htot_ind_int}
	H_{le-m}&=\sum_{i}b_i\left( \Gamma_{i}c_i^\dagger + 
	\Gamma_{i}^\ast c_i\right)\;,
\end{align}
where $b_i$ is the operator describing coupling of the $i$-th excitation state to the low-energy matter excitations with coupling strength $\Gamma_i$.

The mediator modes, the vacuum EM field, and their mutual interaction are described by a Hamiltonian $H_{mp} = H_{m}+H_{p}+H_{m-p}$ that can be compactly represented in matrix form,
\begin{equation}\label{eq:Hmp}
H_{mp} = \begin{pmatrix} \mathbf{c}^\dagger & \mathbf{a}^\dagger \end{pmatrix} \mathbf{H}_{mp}
\begin{pmatrix}  \mathbf{c} \\ \mathbf{a} \end{pmatrix}, \ \
\mathbf{H}_{mp}= \begin{pmatrix}
\boldsymbol{\Omega}& \boldsymbol{\zeta} \\
\boldsymbol{\zeta}^\dagger& \boldsymbol{\omega} \\
\end{pmatrix}.
\end{equation}
It can be diagonalized by a unitary transformation $U$, giving rise to a \emph{polaritonic} 
Hamiltonian describing the \emph{dressed} EM field,
\begin{equation}
H_\mathrm{pol} = U^\dagger H_{mp} U = \sum_n\hbar\tilde{\omega}_n\pi_n^\dagger\pi_n \label{eq:H_pol}\;,
\end{equation}
with $\pi_n^\dagger$ and $\pi_n$ being the creation and annihilation operators of the $n$-th polaritonic mode with frequency $\tilde{\omega}_n$, respectively.
By using the matrix representation of $U$, $\mathbf{U}=(\mathbf{C}, \mathbf{A})$, the total Hamiltonian can be written in the polariton basis as
\begin{equation}\label{eq:Htot_ind_pol}
	{H} =H_{le}+\sum_l\hbar\tilde{\omega}_n\pi_n^\dagger\pi_n+ \sum_{i,n}b_i (\Gamma_{i} C_{in}^\ast \pi_{n}^\dagger+\Gamma_{i}^\ast C_{in} \pi_{n})\;,
\end{equation}
where $C_{in}$ are the coefficients of matrix $\mathbf{C}$. Notice that this Hamiltonian is mathematically equivalent to \autoref{eq:Htot}. This is not unexpected, as the medium-assisted polaritonic operators $\mathbf{f}_p(\mathbf{r}^\prime, \omega)$ themselves arise from the diagonalization of the bare EM modes coupled to bosonic modes representing the cavity material and are thus also polaritonic modes~\cite{Wubs2004,Philbin2010}, but generalizes the derivation from considering only dipole--electric-field interactions to the case of arbitrary interactions between the material (mediator) modes and the low-energy excitations.
Consequently, following a similar procedure for tracing out the 
polaritonic modes~\cite{roman-roche_effective_2022}, we can derive an effective Hamiltonian similar to \autoref{eq:Heff}
\begin{align}\label{eq:Heff_ind}
	{H} _\mathrm{eff}&=H_{le}-\sum_{i,j} b_{i} \Re(\xi_{ij}) b_j \;,
\end{align}
with $\xi_{ij}= \Gamma_i^\ast D_{ij}\Gamma_j$ and  $D_{ij} =\sum_{n}C_{in}C_{jn}^\ast/(\hbar\tilde{\omega}_n)$.
This Hamiltonian shows that an effective interaction between matter operators $b_i$ and $b_j$ is also induced by the coupling with the polaritonic environment and is quantified by strength $\xi_{ij}$.

Since $U$ diagonalizes $H_{mp}$, then $\mathbf{H}_{mp}^{-1}=\mathbf{U}{\boldsymbol{\tilde{\omega}}}^{-1}\mathbf{U}^\dagger$ with $\boldsymbol{\tilde{\omega}}=\mathrm{diag}({\hbar\tilde{\omega}_1,\hbar\tilde{\omega}_2,\ldots})$. 
Consequently, $D_{ij}=\big(\mathbf{C}\boldsymbol{\tilde{\omega}}^{-1}\mathbf{C}^\dagger\big)_{ij}=\big(\mathbf{H}_{mp}^{-1}\big)_{ij}$.
With the help of \autoref{eq:Hmp} and keeping only the block of the matrix that relates the mediator's operators, we obtain
$D_{ij}=\big[(\boldsymbol{\Omega}-\boldsymbol{\zeta}\boldsymbol{\omega}^{-1}\boldsymbol{\zeta}^\dagger)^{-1}\big]_{ij}$.
Assuming that the coupling $\zeta$ is much smaller than the frequency of the mediator $\Omega$ (consistent with the use of the rotating-wave approximation in \autoref{eq:mediator-em-interaction}), we obtain the following expression for the effective coupling
\begin{align}\label{eq:delta_long}
\xi_{ij} &=\dfrac{\Gamma_i^\ast \Gamma_j}{\hbar\Omega} \delta_{ij}+\dfrac{\Gamma_i^\ast\Gamma_j}{\hbar^2\Omega^2}\sum_{n}\dfrac{\boldsymbol{\nu}_i^\ast\cdot\mathbf{E}_n(\mathbf{r}_i) \mathbf{E}_n^\ast(\mathbf{r}_j) \cdot \boldsymbol{\nu}_j}{\hbar\omega_n}\;.
\end{align}
In a similar manner as for the case of direct matter-photon interaction, the sum over EM modes can be performed explicitly, leading to a relation between the effective coupling and the electrostatic ($\omega=0$) Green's function
\begin{align}\label{eq:xi}
\xi_{ij} &=\dfrac{\Gamma_i^\ast \Gamma_j}{\hbar\Omega} \delta_{ij}+\dfrac{\Gamma_i^\ast\Gamma_j}{2\epsilon_0 c_0^2\hbar^2\Omega^2}\left[\omega^2\boldsymbol{\nu}_i^\ast\cdot\mathbf{G}(\mathbf{r}_i,\mathbf{r}_j,\omega)\cdot\boldsymbol{\nu}_j\right]_{\omega=0}\;.
\end{align}

This last expression shows that even though the local mediator modes can be resonant with and strongly coupled to one of the cavity modes, 
the effective matter-matter coupling induced by vacuum EM fields does not depend resonantly on the energies of the mediator and photon modes, but bears an electrostatic nature, as in the case of direct matter-photon interaction.  

\textit{Discussion}.---
Our analysis above culminating in Eqs.~\eqref{eq:lambda_gf} and~\eqref{eq:xi} provides a powerful theoretical shortcut to evaluating the matter-matter interactions mediated by the EM modes under rather general conditions. Therefore, it allows us to understand the potential role of cavity QED in modifying matter excitations by inducing mutual interactions between them. It also places limits on the types of effects that can be achieved with off-resonant interactions where all relevant cavity modes are at significantly higher frequencies than the matter excitations of interest. A corollary of our finding that the resulting matter-matter interaction is electrostatic in nature is that the renormalization of single-particle or single-mode properties by the EM modes could be playing a more important role in the control of matter via the cavity modes.
As noted above, while we have worked in the dipole approximation for simplicity, it is interesting to examine what would change for extended matter excitations, such as conduction electrons. This corresponds to replacing the point-dipole interaction in \autoref{eq:Htot} by an integral over the polarization density~\cite{buhmann_dispersion_2012}, which importantly still corresponds to a linear interaction and thus carries through the remaining derivation unchanged. This would thus lead to a corresponding (double) integral in the final interaction \autoref{eq:lambda_gf}, but would not otherwise modify our observations.
We further note that the above theory could also be applied when other long-range bosonic modes, such as phonons supported by solid state systems, mediate the interactions.

In conclusion, we have studied two different strong light-matter interaction schemes in which cavity-induced modifications of low-energy matter-matter interactions have been predicted, one 
in which the light-matter interaction is direct and another in which a mediator is coupled to both matter and light components, leading to an indirect light-matter interaction. 
In both cases we have found that, by tracing out the photonic degrees of freedom, these modifications can be captured by a reduced Hamiltonian that only involves matter states. An effective matter-matter interaction can thus be induced by the vacuum EM fields associated with a cavity (or any arbitrary material structure in general). However, this effective interaction has a non-resonant character, implying that results with a reduced number of cavity modes can lead to incorrect conclusions regarding the strength and collective properties of the effective interaction. Moreover, by using a macroscopic QED formalism, we have been able to account for the whole spectrum of cavity EM modes, demonstrating that this effective 
matter-matter interaction mediated by vacuum EM fields only depends on the EM response at zero frequency, pointing to an electrostatic origin for this type of cavity-induced modifications of material properties. Our findings provide both evidence for the fundamental understanding of manipulation of matter states via vacuum EM fields and insight for the design of photonic structures that could exhibit large cavity-induced low-energy matter-matter interactions. This could be of interest for the design of novel cavity-modified materials, such as superconductors, where the effective interaction between electrons is a key ingredient for the emergence of superconductivity, or for magnetic materials where long-range order is established due to spin-spin interactions.

\begin{acknowledgments}
\textit{Acknowledgments}.---We thank Diego Fernández de la Pradilla for helpful discussions.
We acknowledge financial support by the Spanish Ministry for Science and Innovation-Agencia Estatal de Investigación (AEI) through Grants PID2021-125894NB-I00, RYC2021-031063-I, and CEX2018-000805-M (through the María de Maeztu program for Units of Excellence in R\&D), and by the European Research Council through Grant No. ERC-2016-StG-714870.
F.J.G.-V. acknowledges financial support by the ``(MAD2D-CM)-UAM7'' project funded by the Comunidad de Madrid, by the Recovery, Transformation and Resilience Plan from Spain, and by NextGenerationEU from the European Union.
\end{acknowledgments}


\begin{thebibliography}{31}%
\makeatletter
\providecommand \@ifxundefined [1]{%
 \@ifx{#1\undefined}
}%
\providecommand \@ifnum [1]{%
 \ifnum #1\expandafter \@firstoftwo
 \else \expandafter \@secondoftwo
 \fi
}%
\providecommand \@ifx [1]{%
 \ifx #1\expandafter \@firstoftwo
 \else \expandafter \@secondoftwo
 \fi
}%
\providecommand \natexlab [1]{#1}%
\providecommand \enquote  [1]{``#1''}%
\providecommand \bibnamefont  [1]{#1}%
\providecommand \bibfnamefont [1]{#1}%
\providecommand \citenamefont [1]{#1}%
\providecommand \href@noop [0]{\@secondoftwo}%
\providecommand \href [0]{\begingroup \@sanitize@url \@href}%
\providecommand \@href[1]{\@@startlink{#1}\@@href}%
\providecommand \@@href[1]{\endgroup#1\@@endlink}%
\providecommand \@sanitize@url [0]{\catcode `\\12\catcode `\$12\catcode
  `\&12\catcode `\#12\catcode `\^12\catcode `\_12\catcode `\%12\relax}%
\providecommand \@@startlink[1]{}%
\providecommand \@@endlink[0]{}%
\providecommand \url  [0]{\begingroup\@sanitize@url \@url }%
\providecommand \@url [1]{\endgroup\@href {#1}{\urlprefix }}%
\providecommand \urlprefix  [0]{URL }%
\providecommand \Eprint [0]{\href }%
\providecommand \doibase [0]{https://doi.org/}%
\providecommand \selectlanguage [0]{\@gobble}%
\providecommand \bibinfo  [0]{\@secondoftwo}%
\providecommand \bibfield  [0]{\@secondoftwo}%
\providecommand \translation [1]{[#1]}%
\providecommand \BibitemOpen [0]{}%
\providecommand \bibitemStop [0]{}%
\providecommand \bibitemNoStop [0]{.\EOS\space}%
\providecommand \EOS [0]{\spacefactor3000\relax}%
\providecommand \BibitemShut  [1]{\csname bibitem#1\endcsname}%
\let\auto@bib@innerbib\@empty
\bibitem [{\citenamefont {Garcia-Vidal}\ \emph {et~al.}(2021)\citenamefont
  {Garcia-Vidal}, \citenamefont {Ciuti},\ and\ \citenamefont
  {Ebbesen}}]{garcia-vidal_manipulating_2021}%
  \BibitemOpen
  \bibfield  {author} {\bibinfo {author} {\bibfnamefont {F.~J.}\ \bibnamefont
  {Garcia-Vidal}}, \bibinfo {author} {\bibfnamefont {C.}~\bibnamefont
  {Ciuti}},\ and\ \bibinfo {author} {\bibfnamefont {T.~W.}\ \bibnamefont
  {Ebbesen}},\ }\bibfield  {title} {\bibinfo {title} {Manipulating matter by
  strong coupling to vacuum fields},\ }\href
  {https://doi.org/10.1126/science.abd0336} {\bibfield  {journal} {\bibinfo
  {journal} {Science}\ }\textbf {\bibinfo {volume} {373}},\ \bibinfo {pages}
  {eabd0336} (\bibinfo {year} {2021})}\BibitemShut {NoStop}%
\bibitem [{\citenamefont {Schlawin}\ \emph {et~al.}(2022)\citenamefont
  {Schlawin}, \citenamefont {Kennes},\ and\ \citenamefont
  {Sentef}}]{schlawin_cavity_2022}%
  \BibitemOpen
  \bibfield  {author} {\bibinfo {author} {\bibfnamefont {F.}~\bibnamefont
  {Schlawin}}, \bibinfo {author} {\bibfnamefont {D.~M.}\ \bibnamefont
  {Kennes}},\ and\ \bibinfo {author} {\bibfnamefont {M.~A.}\ \bibnamefont
  {Sentef}},\ }\bibfield  {title} {\bibinfo {title} {Cavity quantum
  materials},\ }\href {https://doi.org/10.1063/5.0083825} {\bibfield  {journal}
  {\bibinfo  {journal} {Applied Physics Reviews}\ }\textbf {\bibinfo {volume}
  {9}},\ \bibinfo {pages} {011312} (\bibinfo {year} {2022})}\BibitemShut
  {NoStop}%
\bibitem [{\citenamefont {Orgiu}\ \emph {et~al.}(2015)\citenamefont {Orgiu},
  \citenamefont {George}, \citenamefont {Hutchison}, \citenamefont {Devaux},
  \citenamefont {Dayen}, \citenamefont {Doudin}, \citenamefont {Stellacci},
  \citenamefont {Genet}, \citenamefont {Schachenmayer}, \citenamefont {Genes},
  \citenamefont {Pupillo}, \citenamefont {Samorì},\ and\ \citenamefont
  {Ebbesen}}]{orgiu_conductivity_2015}%
  \BibitemOpen
  \bibfield  {author} {\bibinfo {author} {\bibfnamefont {E.}~\bibnamefont
  {Orgiu}}, \bibinfo {author} {\bibfnamefont {J.}~\bibnamefont {George}},
  \bibinfo {author} {\bibfnamefont {J.~A.}\ \bibnamefont {Hutchison}}, \bibinfo
  {author} {\bibfnamefont {E.}~\bibnamefont {Devaux}}, \bibinfo {author}
  {\bibfnamefont {J.~F.}\ \bibnamefont {Dayen}}, \bibinfo {author}
  {\bibfnamefont {B.}~\bibnamefont {Doudin}}, \bibinfo {author} {\bibfnamefont
  {F.}~\bibnamefont {Stellacci}}, \bibinfo {author} {\bibfnamefont
  {C.}~\bibnamefont {Genet}}, \bibinfo {author} {\bibfnamefont
  {J.}~\bibnamefont {Schachenmayer}}, \bibinfo {author} {\bibfnamefont
  {C.}~\bibnamefont {Genes}}, \bibinfo {author} {\bibfnamefont
  {G.}~\bibnamefont {Pupillo}}, \bibinfo {author} {\bibfnamefont
  {P.}~\bibnamefont {Samorì}},\ and\ \bibinfo {author} {\bibfnamefont {T.~W.}\
  \bibnamefont {Ebbesen}},\ }\bibfield  {title} {\bibinfo {title} {Conductivity
  in organic semiconductors hybridized with the vacuum field},\ }\href
  {https://doi.org/10.1038/nmat4392} {\bibfield  {journal} {\bibinfo  {journal}
  {Nature Materials}\ }\textbf {\bibinfo {volume} {14}},\ \bibinfo {pages}
  {1123} (\bibinfo {year} {2015})}\BibitemShut {NoStop}%
\bibitem [{\citenamefont {Feist}\ and\ \citenamefont
  {Garcia-Vidal}(2015)}]{feist_extraordinary_2015}%
  \BibitemOpen
  \bibfield  {author} {\bibinfo {author} {\bibfnamefont {J.}~\bibnamefont
  {Feist}}\ and\ \bibinfo {author} {\bibfnamefont {F.~J.}\ \bibnamefont
  {Garcia-Vidal}},\ }\bibfield  {title} {\bibinfo {title} {Extraordinary
  {Exciton} {Conductance} {Induced} by {Strong} {Coupling}},\ }\href
  {https://doi.org/10.1103/PhysRevLett.114.196402} {\bibfield  {journal}
  {\bibinfo  {journal} {Physical Review Letters}\ }\textbf {\bibinfo {volume}
  {114}},\ \bibinfo {pages} {196402} (\bibinfo {year} {2015})}\BibitemShut
  {NoStop}%
\bibitem [{\citenamefont {Schachenmayer}\ \emph {et~al.}(2015)\citenamefont
  {Schachenmayer}, \citenamefont {Genes}, \citenamefont {Tignone},\ and\
  \citenamefont {Pupillo}}]{schachenmayer_cavity-enhanced_2015}%
  \BibitemOpen
  \bibfield  {author} {\bibinfo {author} {\bibfnamefont {J.}~\bibnamefont
  {Schachenmayer}}, \bibinfo {author} {\bibfnamefont {C.}~\bibnamefont
  {Genes}}, \bibinfo {author} {\bibfnamefont {E.}~\bibnamefont {Tignone}},\
  and\ \bibinfo {author} {\bibfnamefont {G.}~\bibnamefont {Pupillo}},\
  }\bibfield  {title} {\bibinfo {title} {Cavity-{Enhanced} {Transport} of
  {Excitons}},\ }\href {https://doi.org/10.1103/PhysRevLett.114.196403}
  {\bibfield  {journal} {\bibinfo  {journal} {Physical Review Letters}\
  }\textbf {\bibinfo {volume} {114}},\ \bibinfo {pages} {196403} (\bibinfo
  {year} {2015})}\BibitemShut {NoStop}%
\bibitem [{\citenamefont {Paravicini-Bagliani}\ \emph
  {et~al.}(2019)\citenamefont {Paravicini-Bagliani}, \citenamefont
  {Appugliese}, \citenamefont {Richter}, \citenamefont {Valmorra},
  \citenamefont {Keller}, \citenamefont {Beck}, \citenamefont {Bartolo},
  \citenamefont {Rössler}, \citenamefont {Ihn}, \citenamefont {Ensslin},
  \citenamefont {Ciuti}, \citenamefont {Scalari},\ and\ \citenamefont
  {Faist}}]{paravicini-bagliani_magneto-transport_2019}%
  \BibitemOpen
  \bibfield  {author} {\bibinfo {author} {\bibfnamefont {G.~L.}\ \bibnamefont
  {Paravicini-Bagliani}}, \bibinfo {author} {\bibfnamefont {F.}~\bibnamefont
  {Appugliese}}, \bibinfo {author} {\bibfnamefont {E.}~\bibnamefont {Richter}},
  \bibinfo {author} {\bibfnamefont {F.}~\bibnamefont {Valmorra}}, \bibinfo
  {author} {\bibfnamefont {J.}~\bibnamefont {Keller}}, \bibinfo {author}
  {\bibfnamefont {M.}~\bibnamefont {Beck}}, \bibinfo {author} {\bibfnamefont
  {N.}~\bibnamefont {Bartolo}}, \bibinfo {author} {\bibfnamefont
  {C.}~\bibnamefont {Rössler}}, \bibinfo {author} {\bibfnamefont
  {T.}~\bibnamefont {Ihn}}, \bibinfo {author} {\bibfnamefont {K.}~\bibnamefont
  {Ensslin}}, \bibinfo {author} {\bibfnamefont {C.}~\bibnamefont {Ciuti}},
  \bibinfo {author} {\bibfnamefont {G.}~\bibnamefont {Scalari}},\ and\ \bibinfo
  {author} {\bibfnamefont {J.}~\bibnamefont {Faist}},\ }\bibfield  {title}
  {\bibinfo {title} {Magneto-transport controlled by {Landau} polariton
  states},\ }\href {https://doi.org/10.1038/s41567-018-0346-y} {\bibfield
  {journal} {\bibinfo  {journal} {Nature Physics}\ }\textbf {\bibinfo {volume}
  {15}},\ \bibinfo {pages} {186} (\bibinfo {year} {2019})}\BibitemShut
  {NoStop}%
\bibitem [{\citenamefont {Mazza}\ and\ \citenamefont
  {Georges}(2019)}]{mazza_superradiant_2019}%
  \BibitemOpen
  \bibfield  {author} {\bibinfo {author} {\bibfnamefont {G.}~\bibnamefont
  {Mazza}}\ and\ \bibinfo {author} {\bibfnamefont {A.}~\bibnamefont
  {Georges}},\ }\bibfield  {title} {\bibinfo {title} {Superradiant {Quantum}
  {Materials}},\ }\href {https://doi.org/10.1103/PhysRevLett.122.017401}
  {\bibfield  {journal} {\bibinfo  {journal} {Physical Review Letters}\
  }\textbf {\bibinfo {volume} {122}},\ \bibinfo {pages} {017401} (\bibinfo
  {year} {2019})}\BibitemShut {NoStop}%
\bibitem [{\citenamefont {Yokoshi}\ \emph {et~al.}(2017)\citenamefont
  {Yokoshi}, \citenamefont {Odagiri}, \citenamefont {Ishikawa},\ and\
  \citenamefont {Ishihara}}]{Yokoshi2017}%
  \BibitemOpen
  \bibfield  {author} {\bibinfo {author} {\bibfnamefont {N.}~\bibnamefont
  {Yokoshi}}, \bibinfo {author} {\bibfnamefont {K.}~\bibnamefont {Odagiri}},
  \bibinfo {author} {\bibfnamefont {A.}~\bibnamefont {Ishikawa}},\ and\
  \bibinfo {author} {\bibfnamefont {H.}~\bibnamefont {Ishihara}},\ }\bibfield
  {title} {\bibinfo {title} {{Synchronization {{Dynamics}} in a {{Designed Open
  System}}}},\ }\href {https://doi.org/10.1103/PhysRevLett.118.203601}
  {\bibfield  {journal} {\bibinfo  {journal} {Phys. Rev. Lett.}\ }\textbf
  {\bibinfo {volume} {118}},\ \bibinfo {pages} {203601} (\bibinfo {year}
  {2017})}\BibitemShut {NoStop}%
\bibitem [{\citenamefont {Kinoshita}\ \emph {et~al.}(2019)\citenamefont
  {Kinoshita}, \citenamefont {Matsuda}, \citenamefont {Takahashi},
  \citenamefont {Ichimiya}, \citenamefont {Ashida}, \citenamefont {Furukawa},
  \citenamefont {Nakayama},\ and\ \citenamefont {Ishihara}}]{Kinoshita2019}%
  \BibitemOpen
  \bibfield  {author} {\bibinfo {author} {\bibfnamefont {T.}~\bibnamefont
  {Kinoshita}}, \bibinfo {author} {\bibfnamefont {T.}~\bibnamefont {Matsuda}},
  \bibinfo {author} {\bibfnamefont {T.}~\bibnamefont {Takahashi}}, \bibinfo
  {author} {\bibfnamefont {M.}~\bibnamefont {Ichimiya}}, \bibinfo {author}
  {\bibfnamefont {M.}~\bibnamefont {Ashida}}, \bibinfo {author} {\bibfnamefont
  {Y.}~\bibnamefont {Furukawa}}, \bibinfo {author} {\bibfnamefont
  {M.}~\bibnamefont {Nakayama}},\ and\ \bibinfo {author} {\bibfnamefont
  {H.}~\bibnamefont {Ishihara}},\ }\bibfield  {title} {\bibinfo {title}
  {{Synergetic {{Enhancement}} of {{Light-Matter Interaction}} by
  {{Nonlocality}} and {{Band Degeneracy}} in {{ZnO Thin Films}}}},\ }\href
  {https://doi.org/10.1103/PhysRevLett.122.157401} {\bibfield  {journal}
  {\bibinfo  {journal} {Phys. Rev. Lett.}\ }\textbf {\bibinfo {volume} {122}},\
  \bibinfo {pages} {157401} (\bibinfo {year} {2019})}\BibitemShut {NoStop}%
\bibitem [{\citenamefont {Appugliese}\ \emph {et~al.}(2022)\citenamefont
  {Appugliese}, \citenamefont {Enkner}, \citenamefont {Paravicini-Bagliani},
  \citenamefont {Beck}, \citenamefont {Reichl}, \citenamefont {Wegscheider},
  \citenamefont {Scalari}, \citenamefont {Ciuti},\ and\ \citenamefont
  {Faist}}]{appugliese_breakdown_2022}%
  \BibitemOpen
  \bibfield  {author} {\bibinfo {author} {\bibfnamefont {F.}~\bibnamefont
  {Appugliese}}, \bibinfo {author} {\bibfnamefont {J.}~\bibnamefont {Enkner}},
  \bibinfo {author} {\bibfnamefont {G.~L.}\ \bibnamefont
  {Paravicini-Bagliani}}, \bibinfo {author} {\bibfnamefont {M.}~\bibnamefont
  {Beck}}, \bibinfo {author} {\bibfnamefont {C.}~\bibnamefont {Reichl}},
  \bibinfo {author} {\bibfnamefont {W.}~\bibnamefont {Wegscheider}}, \bibinfo
  {author} {\bibfnamefont {G.}~\bibnamefont {Scalari}}, \bibinfo {author}
  {\bibfnamefont {C.}~\bibnamefont {Ciuti}},\ and\ \bibinfo {author}
  {\bibfnamefont {J.}~\bibnamefont {Faist}},\ }\bibfield  {title} {\bibinfo
  {title} {Breakdown of topological protection by cavity vacuum fields in the
  integer quantum {Hall} effect},\ }\href
  {https://doi.org/10.1126/science.abl5818} {\bibfield  {journal} {\bibinfo
  {journal} {Science}\ }\textbf {\bibinfo {volume} {375}},\ \bibinfo {pages}
  {1030} (\bibinfo {year} {2022})}\BibitemShut {NoStop}%
\bibitem [{\citenamefont {Schlawin}\ \emph {et~al.}(2019)\citenamefont
  {Schlawin}, \citenamefont {Cavalleri},\ and\ \citenamefont
  {Jaksch}}]{schlawin_cavity-mediated_2019}%
  \BibitemOpen
  \bibfield  {author} {\bibinfo {author} {\bibfnamefont {F.}~\bibnamefont
  {Schlawin}}, \bibinfo {author} {\bibfnamefont {A.}~\bibnamefont
  {Cavalleri}},\ and\ \bibinfo {author} {\bibfnamefont {D.}~\bibnamefont
  {Jaksch}},\ }\bibfield  {title} {\bibinfo {title} {Cavity-{Mediated}
  {Electron}-{Photon} {Superconductivity}},\ }\href
  {https://doi.org/10.1103/PhysRevLett.122.133602} {\bibfield  {journal}
  {\bibinfo  {journal} {Physical Review Letters}\ }\textbf {\bibinfo {volume}
  {122}},\ \bibinfo {pages} {133602} (\bibinfo {year} {2019})}\BibitemShut
  {NoStop}%
\bibitem [{\citenamefont {Curtis}\ \emph {et~al.}(2019)\citenamefont {Curtis},
  \citenamefont {Raines}, \citenamefont {Allocca}, \citenamefont {Hafezi},\
  and\ \citenamefont {Galitski}}]{curtis_cavity_2019}%
  \BibitemOpen
  \bibfield  {author} {\bibinfo {author} {\bibfnamefont {J.~B.}\ \bibnamefont
  {Curtis}}, \bibinfo {author} {\bibfnamefont {Z.~M.}\ \bibnamefont {Raines}},
  \bibinfo {author} {\bibfnamefont {A.~A.}\ \bibnamefont {Allocca}}, \bibinfo
  {author} {\bibfnamefont {M.}~\bibnamefont {Hafezi}},\ and\ \bibinfo {author}
  {\bibfnamefont {V.~M.}\ \bibnamefont {Galitski}},\ }\bibfield  {title}
  {\bibinfo {title} {Cavity {Quantum} {Eliashberg} {Enhancement} of
  {Superconductivity}},\ }\href
  {https://doi.org/10.1103/PhysRevLett.122.167002} {\bibfield  {journal}
  {\bibinfo  {journal} {Physical Review Letters}\ }\textbf {\bibinfo {volume}
  {122}},\ \bibinfo {pages} {167002} (\bibinfo {year} {2019})}\BibitemShut
  {NoStop}%
\bibitem [{\citenamefont {Roman-Roche}\ \emph {et~al.}(2021)\citenamefont
  {Roman-Roche}, \citenamefont {Luis},\ and\ \citenamefont
  {Zueco}}]{roman-roche_photon_2021}%
  \BibitemOpen
  \bibfield  {author} {\bibinfo {author} {\bibfnamefont {J.}~\bibnamefont
  {Roman-Roche}}, \bibinfo {author} {\bibfnamefont {F.}~\bibnamefont {Luis}},\
  and\ \bibinfo {author} {\bibfnamefont {D.}~\bibnamefont {Zueco}},\ }\bibfield
   {title} {\bibinfo {title} {Photon {Condensation} and {Enhanced} {Magnetism}
  in {Cavity} {QED}},\ }\href {https://doi.org/10.1103/PhysRevLett.127.167201}
  {\bibfield  {journal} {\bibinfo  {journal} {Physical Review Letters}\
  }\textbf {\bibinfo {volume} {127}},\ \bibinfo {pages} {167201} (\bibinfo
  {year} {2021})}\BibitemShut {NoStop}%
\bibitem [{\citenamefont {Ashida}\ \emph {et~al.}(2020)\citenamefont {Ashida},
  \citenamefont {İmamoğlu}, \citenamefont {Faist}, \citenamefont {Jaksch},
  \citenamefont {Cavalleri},\ and\ \citenamefont
  {Demler}}]{ashida_quantum_2020}%
  \BibitemOpen
  \bibfield  {author} {\bibinfo {author} {\bibfnamefont {Y.}~\bibnamefont
  {Ashida}}, \bibinfo {author} {\bibfnamefont {A.}~\bibnamefont {İmamoğlu}},
  \bibinfo {author} {\bibfnamefont {J.}~\bibnamefont {Faist}}, \bibinfo
  {author} {\bibfnamefont {D.}~\bibnamefont {Jaksch}}, \bibinfo {author}
  {\bibfnamefont {A.}~\bibnamefont {Cavalleri}},\ and\ \bibinfo {author}
  {\bibfnamefont {E.}~\bibnamefont {Demler}},\ }\bibfield  {title} {\bibinfo
  {title} {Quantum {Electrodynamic} {Control} of {Matter}: {Cavity}-{Enhanced}
  {Ferroelectric} {Phase} {Transition}},\ }\href
  {https://doi.org/10.1103/PhysRevX.10.041027} {\bibfield  {journal} {\bibinfo
  {journal} {Physical Review X}\ }\textbf {\bibinfo {volume} {10}},\ \bibinfo
  {pages} {041027} (\bibinfo {year} {2020})}\BibitemShut {NoStop}%
\bibitem [{\citenamefont {Lenk}\ \emph {et~al.}(2022)\citenamefont {Lenk},
  \citenamefont {Li}, \citenamefont {Werner},\ and\ \citenamefont
  {Eckstein}}]{lenk_dynamical_2022}%
  \BibitemOpen
  \bibfield  {author} {\bibinfo {author} {\bibfnamefont {K.}~\bibnamefont
  {Lenk}}, \bibinfo {author} {\bibfnamefont {J.}~\bibnamefont {Li}}, \bibinfo
  {author} {\bibfnamefont {P.}~\bibnamefont {Werner}},\ and\ \bibinfo {author}
  {\bibfnamefont {M.}~\bibnamefont {Eckstein}},\ }\bibfield  {title} {\bibinfo
  {title} {Dynamical mean-field study of a photon-mediated ferroelectric phase
  transition},\ }\href {https://doi.org/10.1103/PhysRevB.106.245124} {\bibfield
   {journal} {\bibinfo  {journal} {Physical Review B}\ }\textbf {\bibinfo
  {volume} {106}},\ \bibinfo {pages} {245124} (\bibinfo {year}
  {2022})}\BibitemShut {NoStop}%
\bibitem [{\citenamefont {Sentef}\ \emph {et~al.}(2018)\citenamefont {Sentef},
  \citenamefont {Ruggenthaler},\ and\ \citenamefont
  {Rubio}}]{sentef_cavity_2018}%
  \BibitemOpen
  \bibfield  {author} {\bibinfo {author} {\bibfnamefont {M.~A.}\ \bibnamefont
  {Sentef}}, \bibinfo {author} {\bibfnamefont {M.}~\bibnamefont
  {Ruggenthaler}},\ and\ \bibinfo {author} {\bibfnamefont {A.}~\bibnamefont
  {Rubio}},\ }\bibfield  {title} {\bibinfo {title} {Cavity
  quantum-electrodynamical polaritonically enhanced electron-phonon coupling
  and its influence on superconductivity},\ }\href
  {https://doi.org/10.1126/sciadv.aau6969} {\bibfield  {journal} {\bibinfo
  {journal} {Science Advances}\ }\textbf {\bibinfo {volume} {4}},\ \bibinfo
  {pages} {eaau6969} (\bibinfo {year} {2018})}\BibitemShut {NoStop}%
\bibitem [{\citenamefont {Hagenmüller}\ \emph {et~al.}(2019)\citenamefont
  {Hagenmüller}, \citenamefont {Schachenmayer}, \citenamefont {Genet},
  \citenamefont {Ebbesen},\ and\ \citenamefont
  {Pupillo}}]{hagenmuller_enhancement_2019}%
  \BibitemOpen
  \bibfield  {author} {\bibinfo {author} {\bibfnamefont {D.}~\bibnamefont
  {Hagenmüller}}, \bibinfo {author} {\bibfnamefont {J.}~\bibnamefont
  {Schachenmayer}}, \bibinfo {author} {\bibfnamefont {C.}~\bibnamefont
  {Genet}}, \bibinfo {author} {\bibfnamefont {T.~W.}\ \bibnamefont {Ebbesen}},\
  and\ \bibinfo {author} {\bibfnamefont {G.}~\bibnamefont {Pupillo}},\
  }\bibfield  {title} {\bibinfo {title} {Enhancement of the
  {Electron}–{Phonon} {Scattering} {Induced} by {Intrinsic} {Surface}
  {Plasmon}–{Phonon} {Polaritons}},\ }\href
  {https://doi.org/10.1021/acsphotonics.9b00268} {\bibfield  {journal}
  {\bibinfo  {journal} {ACS Photonics}\ }\textbf {\bibinfo {volume} {6}},\
  \bibinfo {pages} {1073} (\bibinfo {year} {2019})}\BibitemShut {NoStop}%
\bibitem [{\citenamefont {Thomas}\ \emph {et~al.}(2021)\citenamefont {Thomas},
  \citenamefont {Devaux}, \citenamefont {Nagarajan}, \citenamefont {Rogez},
  \citenamefont {Seidel}, \citenamefont {Richard}, \citenamefont {Genet},
  \citenamefont {Drillon},\ and\ \citenamefont {Ebbesen}}]{thomas_large_2021}%
  \BibitemOpen
  \bibfield  {author} {\bibinfo {author} {\bibfnamefont {A.}~\bibnamefont
  {Thomas}}, \bibinfo {author} {\bibfnamefont {E.}~\bibnamefont {Devaux}},
  \bibinfo {author} {\bibfnamefont {K.}~\bibnamefont {Nagarajan}}, \bibinfo
  {author} {\bibfnamefont {G.}~\bibnamefont {Rogez}}, \bibinfo {author}
  {\bibfnamefont {M.}~\bibnamefont {Seidel}}, \bibinfo {author} {\bibfnamefont
  {F.}~\bibnamefont {Richard}}, \bibinfo {author} {\bibfnamefont
  {C.}~\bibnamefont {Genet}}, \bibinfo {author} {\bibfnamefont
  {M.}~\bibnamefont {Drillon}},\ and\ \bibinfo {author} {\bibfnamefont {T.~W.}\
  \bibnamefont {Ebbesen}},\ }\bibfield  {title} {\bibinfo {title} {Large
  {Enhancement} of {Ferromagnetism} under a {Collective} {Strong} {Coupling} of
  {YBCO} {Nanoparticles}},\ }\href
  {https://doi.org/10.1021/acs.nanolett.1c00973} {\bibfield  {journal}
  {\bibinfo  {journal} {Nano Letters}\ }\textbf {\bibinfo {volume} {21}},\
  \bibinfo {pages} {4365} (\bibinfo {year} {2021})}\BibitemShut {NoStop}%
\bibitem [{\citenamefont {Curtis}\ \emph {et~al.}(2022)\citenamefont {Curtis},
  \citenamefont {Grankin}, \citenamefont {Poniatowski}, \citenamefont
  {Galitski}, \citenamefont {Narang},\ and\ \citenamefont
  {Demler}}]{curtis_cavity_2022}%
  \BibitemOpen
  \bibfield  {author} {\bibinfo {author} {\bibfnamefont {J.~B.}\ \bibnamefont
  {Curtis}}, \bibinfo {author} {\bibfnamefont {A.}~\bibnamefont {Grankin}},
  \bibinfo {author} {\bibfnamefont {N.~R.}\ \bibnamefont {Poniatowski}},
  \bibinfo {author} {\bibfnamefont {V.~M.}\ \bibnamefont {Galitski}}, \bibinfo
  {author} {\bibfnamefont {P.}~\bibnamefont {Narang}},\ and\ \bibinfo {author}
  {\bibfnamefont {E.}~\bibnamefont {Demler}},\ }\bibfield  {title} {\bibinfo
  {title} {Cavity magnon-polaritons in cuprate parent compounds},\ }\href
  {https://doi.org/10.1103/PhysRevResearch.4.013101} {\bibfield  {journal}
  {\bibinfo  {journal} {Physical Review Research}\ }\textbf {\bibinfo {volume}
  {4}},\ \bibinfo {pages} {013101} (\bibinfo {year} {2022})}\BibitemShut
  {NoStop}%
\bibitem [{\citenamefont {Wang}\ and\ \citenamefont
  {Hioe}(1973)}]{wang_phase_1973}%
  \BibitemOpen
  \bibfield  {author} {\bibinfo {author} {\bibfnamefont {Y.~K.}\ \bibnamefont
  {Wang}}\ and\ \bibinfo {author} {\bibfnamefont {F.~T.}\ \bibnamefont
  {Hioe}},\ }\bibfield  {title} {\bibinfo {title} {Phase {Transition} in the
  {Dicke} {Model} of {Superradiance}},\ }\href
  {https://doi.org/10.1103/PhysRevA.7.831} {\bibfield  {journal} {\bibinfo
  {journal} {Physical Review A}\ }\textbf {\bibinfo {volume} {7}},\ \bibinfo
  {pages} {831} (\bibinfo {year} {1973})}\BibitemShut {NoStop}%
\bibitem [{\citenamefont {Hepp}\ and\ \citenamefont
  {Lieb}(1973)}]{hepp_superradiant_1973}%
  \BibitemOpen
  \bibfield  {author} {\bibinfo {author} {\bibfnamefont {K.}~\bibnamefont
  {Hepp}}\ and\ \bibinfo {author} {\bibfnamefont {E.~H.}\ \bibnamefont
  {Lieb}},\ }\bibfield  {title} {\bibinfo {title} {On the superradiant phase
  transition for molecules in a quantized radiation field: the dicke maser
  model},\ }\href {https://doi.org/10.1016/0003-4916(73)90039-0} {\bibfield
  {journal} {\bibinfo  {journal} {Annals of Physics}\ }\textbf {\bibinfo
  {volume} {76}},\ \bibinfo {pages} {360} (\bibinfo {year} {1973})}\BibitemShut
  {NoStop}%
\bibitem [{\citenamefont {Coleman}(2015)}]{coleman_introduction_2015}%
  \BibitemOpen
  \bibfield  {author} {\bibinfo {author} {\bibfnamefont {P.}~\bibnamefont
  {Coleman}},\ }\href {https://doi.org/10.1017/CBO9781139020916} {\emph
  {\bibinfo {title} {Introduction to {Many}-{Body} {Physics}}}}\ (\bibinfo
  {publisher} {Cambridge University Press},\ \bibinfo {address} {Cambridge},\
  \bibinfo {year} {2015})\BibitemShut {NoStop}%
\bibitem [{\citenamefont {Román-Roche}\ and\ \citenamefont
  {Zueco}(2022)}]{roman-roche_effective_2022}%
  \BibitemOpen
  \bibfield  {author} {\bibinfo {author} {\bibfnamefont {J.}~\bibnamefont
  {Román-Roche}}\ and\ \bibinfo {author} {\bibfnamefont {D.}~\bibnamefont
  {Zueco}},\ }\bibfield  {title} {\bibinfo {title} {Effective theory for matter
  in non-perturbative cavity {QED}},\ }\href
  {https://doi.org/10.21468/SciPostPhysLectNotes.50} {\bibfield  {journal}
  {\bibinfo  {journal} {SciPost Physics Lecture Notes}\ ,\ \bibinfo {pages}
  {50}} (\bibinfo {year} {2022})}\BibitemShut {NoStop}%
\bibitem [{sup()}]{supplemental}%
  \BibitemOpen
  \href@noop {} {\bibinfo {title} {See {S}upplemental {M}aterial for
  the details on the derivation of the effective Hamiltonian, on the
  calculation of the effective coupling for the case of a continuum of EM
  modes, and for the discussion of the free-space case and cutoff dependence of
  the integral over frequencies.}}\BibitemShut {Stop}%
\bibitem [{\citenamefont {Buhmann}(2012)}]{buhmann_dispersion_2012}%
  \BibitemOpen
  \bibfield  {author} {\bibinfo {author} {\bibfnamefont {S.~Y.}\ \bibnamefont
  {Buhmann}},\ }\href {https://doi.org/10.1007/978-3-642-32484-0} {\emph
  {\bibinfo {title} {Dispersion {Forces} {I}}}},\ \bibinfo {series} {Springer
  {Tracts} in {Modern} {Physics}}, Vol.\ \bibinfo {volume} {247}\ (\bibinfo
  {publisher} {Springer Berlin Heidelberg},\ \bibinfo {address} {Berlin,
  Heidelberg},\ \bibinfo {year} {2012})\BibitemShut {NoStop}%
\bibitem [{\citenamefont {Novotny}\ and\ \citenamefont
  {Hecht}(2012)}]{novotny_principles_2012}%
  \BibitemOpen
  \bibfield  {author} {\bibinfo {author} {\bibfnamefont {L.}~\bibnamefont
  {Novotny}}\ and\ \bibinfo {author} {\bibfnamefont {B.}~\bibnamefont
  {Hecht}},\ }\href {https://doi.org/10.1017/CBO9780511794193} {\emph {\bibinfo
  {title} {Principles of {Nano}-{Optics}}}},\ \bibinfo {edition} {2nd}\ ed.\
  (\bibinfo  {publisher} {Cambridge University Press},\ \bibinfo {address}
  {Cambridge},\ \bibinfo {year} {2012})\BibitemShut {NoStop}%
\bibitem [{\citenamefont {Casimir}\ and\ \citenamefont
  {Polder}(1948)}]{Casimir1948}%
  \BibitemOpen
  \bibfield  {author} {\bibinfo {author} {\bibfnamefont {H.~B.~G.}\
  \bibnamefont {Casimir}}\ and\ \bibinfo {author} {\bibfnamefont
  {D.}~\bibnamefont {Polder}},\ }\bibfield  {title} {\bibinfo {title} {{The
  {{Influence}} of {{Retardation}} on the {{London-van}} Der {{Waals
  Forces}}}},\ }\href {https://doi.org/10.1103/PhysRev.73.360} {\bibfield
  {journal} {\bibinfo  {journal} {Phys. Rev.}\ }\textbf {\bibinfo {volume}
  {73}},\ \bibinfo {pages} {360} (\bibinfo {year} {1948})}\BibitemShut
  {NoStop}%
\bibitem [{\citenamefont {De~Bernardis}\ \emph {et~al.}(2018)\citenamefont
  {De~Bernardis}, \citenamefont {Jaako},\ and\ \citenamefont
  {Rabl}}]{de_bernardis_cavity_2018}%
  \BibitemOpen
  \bibfield  {author} {\bibinfo {author} {\bibfnamefont {D.}~\bibnamefont
  {De~Bernardis}}, \bibinfo {author} {\bibfnamefont {T.}~\bibnamefont
  {Jaako}},\ and\ \bibinfo {author} {\bibfnamefont {P.}~\bibnamefont {Rabl}},\
  }\bibfield  {title} {\bibinfo {title} {Cavity quantum electrodynamics in the
  nonperturbative regime},\ }\href {https://doi.org/10.1103/PhysRevA.97.043820}
  {\bibfield  {journal} {\bibinfo  {journal} {Physical Review A}\ }\textbf
  {\bibinfo {volume} {97}},\ \bibinfo {pages} {043820} (\bibinfo {year}
  {2018})}\BibitemShut {NoStop}%
\bibitem [{\citenamefont {S\'aez-Bl\'azquez}\ \emph {et~al.}(2023)\citenamefont
  {S\'aez-Bl\'azquez}, \citenamefont {de~Bernardis}, \citenamefont {Feist},\
  and\ \citenamefont {Rabl}}]{Saez-Blazquez2023}%
  \BibitemOpen
  \bibfield  {author} {\bibinfo {author} {\bibfnamefont {R.}~\bibnamefont
  {S\'aez-Bl\'azquez}}, \bibinfo {author} {\bibfnamefont {D.}~\bibnamefont
  {de~Bernardis}}, \bibinfo {author} {\bibfnamefont {J.}~\bibnamefont
  {Feist}},\ and\ \bibinfo {author} {\bibfnamefont {P.}~\bibnamefont {Rabl}},\
  }\bibfield  {title} {\bibinfo {title} {Can we observe nonperturbative vacuum
  shifts in cavity qed?},\ }\href
  {https://doi.org/10.1103/PhysRevLett.131.013602} {\bibfield  {journal}
  {\bibinfo  {journal} {Phys. Rev. Lett.}\ }\textbf {\bibinfo {volume} {131}},\
  \bibinfo {pages} {013602} (\bibinfo {year} {2023})}\BibitemShut {NoStop}%
\bibitem [{\citenamefont {Wubs}\ \emph {et~al.}(2004)\citenamefont {Wubs},
  \citenamefont {Suttorp},\ and\ \citenamefont {Lagendijk}}]{Wubs2004}%
  \BibitemOpen
  \bibfield  {author} {\bibinfo {author} {\bibfnamefont {M.}~\bibnamefont
  {Wubs}}, \bibinfo {author} {\bibfnamefont {L.~G.}\ \bibnamefont {Suttorp}},\
  and\ \bibinfo {author} {\bibfnamefont {A.}~\bibnamefont {Lagendijk}},\
  }\bibfield  {title} {\bibinfo {title} {{Multiple-Scattering Approach to
  Interatomic Interactions and Superradiance in Inhomogeneous Dielectrics}},\
  }\href {https://doi.org/10.1103/PhysRevA.70.053823} {\bibfield  {journal}
  {\bibinfo  {journal} {Phys. Rev. A}\ }\textbf {\bibinfo {volume} {70}},\
  \bibinfo {pages} {053823} (\bibinfo {year} {2004})}\BibitemShut {NoStop}%
\bibitem [{\citenamefont {Philbin}(2010)}]{Philbin2010}%
  \BibitemOpen
  \bibfield  {author} {\bibinfo {author} {\bibfnamefont {T.~G.}\ \bibnamefont
  {Philbin}},\ }\bibfield  {title} {\bibinfo {title} {{Canonical Quantization
  of Macroscopic Electromagnetism}},\ }\href
  {https://doi.org/10.1088/1367-2630/12/12/123008} {\bibfield  {journal}
  {\bibinfo  {journal} {New J. Phys.}\ }\textbf {\bibinfo {volume} {12}},\
  \bibinfo {pages} {123008} (\bibinfo {year} {2010})}\BibitemShut {NoStop}%
\end{thebibliography}

\begin{thebibliography}{5}%
\makeatletter
\providecommand \@ifxundefined [1]{%
 \@ifx{#1\undefined}
}%
\providecommand \@ifnum [1]{%
 \ifnum #1\expandafter \@firstoftwo
 \else \expandafter \@secondoftwo
 \fi
}%
\providecommand \@ifx [1]{%
 \ifx #1\expandafter \@firstoftwo
 \else \expandafter \@secondoftwo
 \fi
}%
\providecommand \natexlab [1]{#1}%
\providecommand \enquote  [1]{``#1''}%
\providecommand \bibnamefont  [1]{#1}%
\providecommand \bibfnamefont [1]{#1}%
\providecommand \citenamefont [1]{#1}%
\providecommand \href@noop [0]{\@secondoftwo}%
\providecommand \href [0]{\begingroup \@sanitize@url \@href}%
\providecommand \@href[1]{\@@startlink{#1}\@@href}%
\providecommand \@@href[1]{\endgroup#1\@@endlink}%
\providecommand \@sanitize@url [0]{\catcode `\\12\catcode `\$12\catcode
  `\&12\catcode `\#12\catcode `\^12\catcode `\_12\catcode `\%12\relax}%
\providecommand \@@startlink[1]{}%
\providecommand \@@endlink[0]{}%
\providecommand \url  [0]{\begingroup\@sanitize@url \@url }%
\providecommand \@url [1]{\endgroup\@href {#1}{\urlprefix }}%
\providecommand \urlprefix  [0]{URL }%
\providecommand \Eprint [0]{\href }%
\providecommand \doibase [0]{https://doi.org/}%
\providecommand \selectlanguage [0]{\@gobble}%
\providecommand \bibinfo  [0]{\@secondoftwo}%
\providecommand \bibfield  [0]{\@secondoftwo}%
\providecommand \translation [1]{[#1]}%
\providecommand \BibitemOpen [0]{}%
\providecommand \bibitemStop [0]{}%
\providecommand \bibitemNoStop [0]{.\EOS\space}%
\providecommand \EOS [0]{\spacefactor3000\relax}%
\providecommand \BibitemShut  [1]{\csname bibitem#1\endcsname}%
\let\auto@bib@innerbib\@empty
\bibitem [{\citenamefont {Wang}\ and\ \citenamefont
  {Hioe}(1973)}]{wang_phase_1973_SM}%
  \BibitemOpen
  \bibfield  {author} {\bibinfo {author} {\bibfnamefont {Y.~K.}\ \bibnamefont
  {Wang}}\ and\ \bibinfo {author} {\bibfnamefont {F.~T.}\ \bibnamefont
  {Hioe}},\ }\bibfield  {title} {\bibinfo {title} {Phase {Transition} in the
  {Dicke} {Model} of {Superradiance}},\ }\href
  {https://doi.org/10.1103/PhysRevA.7.831} {\bibfield  {journal} {\bibinfo
  {journal} {Physical Review A}\ }\textbf {\bibinfo {volume} {7}},\ \bibinfo
  {pages} {831} (\bibinfo {year} {1973})}\BibitemShut {NoStop}%
\bibitem [{\citenamefont {Hepp}\ and\ \citenamefont
  {Lieb}(1973)}]{hepp_superradiant_1973_SM}%
  \BibitemOpen
  \bibfield  {author} {\bibinfo {author} {\bibfnamefont {K.}~\bibnamefont
  {Hepp}}\ and\ \bibinfo {author} {\bibfnamefont {E.~H.}\ \bibnamefont
  {Lieb}},\ }\bibfield  {title} {\bibinfo {title} {On the superradiant phase
  transition for molecules in a quantized radiation field: the dicke maser
  model},\ }\href {https://doi.org/10.1016/0003-4916(73)90039-0} {\bibfield
  {journal} {\bibinfo  {journal} {Annals of Physics}\ }\textbf {\bibinfo
  {volume} {76}},\ \bibinfo {pages} {360} (\bibinfo {year} {1973})}\BibitemShut
  {NoStop}%
\bibitem [{\citenamefont {Coleman}(2015)}]{coleman_introduction_2015_SM}%
  \BibitemOpen
  \bibfield  {author} {\bibinfo {author} {\bibfnamefont {P.}~\bibnamefont
  {Coleman}},\ }\href {https://doi.org/10.1017/CBO9781139020916} {\emph
  {\bibinfo {title} {Introduction to {Many}-{Body} {Physics}}}}\ (\bibinfo
  {publisher} {Cambridge University Press},\ \bibinfo {address} {Cambridge},\
  \bibinfo {year} {2015})\BibitemShut {NoStop}%
\bibitem [{\citenamefont {Román-Roche}\ and\ \citenamefont
  {Zueco}(2022)}]{roman-roche_effective_2022_SM}%
  \BibitemOpen
  \bibfield  {author} {\bibinfo {author} {\bibfnamefont {J.}~\bibnamefont
  {Román-Roche}}\ and\ \bibinfo {author} {\bibfnamefont {D.}~\bibnamefont
  {Zueco}},\ }\bibfield  {title} {\bibinfo {title} {Effective theory for matter
  in non-perturbative cavity {QED}},\ }\href
  {https://doi.org/10.21468/SciPostPhysLectNotes.50} {\bibfield  {journal}
  {\bibinfo  {journal} {SciPost Physics Lecture Notes}\ ,\ \bibinfo {pages}
  {50}} (\bibinfo {year} {2022})}\BibitemShut {NoStop}%
\bibitem [{\citenamefont {Buhmann}(2012)}]{buhmann_dispersion_2012_SM}%
  \BibitemOpen
  \bibfield  {author} {\bibinfo {author} {\bibfnamefont {S.~Y.}\ \bibnamefont
  {Buhmann}},\ }\href {https://doi.org/10.1007/978-3-642-32484-0} {\emph
  {\bibinfo {title} {Dispersion {Forces} {I}}}},\ \bibinfo {series} {Springer
  {Tracts} in {Modern} {Physics}}, Vol.\ \bibinfo {volume} {247}\ (\bibinfo
  {publisher} {Springer Berlin Heidelberg},\ \bibinfo {address} {Berlin,
  Heidelberg},\ \bibinfo {year} {2012})\BibitemShut {NoStop}%
\end{thebibliography}
%

\pagebreak
\widetext
\begin{center}
	\textbf{\large Supplemental Material: Electrostatic nature of cavity-mediated\\interactions between low-energy matter excitations}
\end{center}
\setcounter{equation}{0}
\setcounter{figure}{0}
\setcounter{table}{0}
\setcounter{page}{1}
\makeatletter
\renewcommand{\theequation}{S\arabic{equation}}
\renewcommand{\thefigure}{S\arabic{figure}}
\renewcommand{\bibnumfmt}[1]{[S#1]}
\renewcommand{\citenumfont}[1]{S#1}

\section*{Derivation of the Effective Hamiltonian}\label{sec:Heff_SM}
We are interested in deriving a Hamiltonian able to encapsulate the effect of the electromagnetic environment on the matter states. 
In other words, our interest lies on deriving an effective matter-matter Hamiltonian. 
To begin with, we consider the single-mode light-matter Hamiltonian 
\begin{equation}\label{eq:Htot_SM}
	H=H_{le}+\hbar\omega a^\dagger a +\sum_{i}\widehat{\boldsymbol{\mu}}_i\cdot\left[\mathbf{E}(\mathbf{r}_i)a+\mathbf{E}^\ast(\mathbf{r}_i)a^\dagger\right]\;.
\end{equation}
The effective matter-matter interactions can be obtained by tracing out the photonic environment~\cite{wang_phase_1973_SM,hepp_superradiant_1973_SM} and more specifically by employing the path integral framework as detailed in~\cite{coleman_introduction_2015_SM,roman-roche_effective_2022_SM}.

The canonical partition function, $Z=\mathrm{Tr}[\exp(-\beta H)]$ with $\beta = 1/(k_B T)$ and $k_B$ the Boltzmann constant, can be used for studying material properties. 
In systems where light interacts with matter, the photonic coherent state is a convenient basis for calculating the partition function. 
The coherent state $\ket{\alpha}$ is eigenstate of the annihilation operator $a$,  $a\ket{\alpha}=\alpha\ket{\alpha}$, has complex eigenvalues, $\alpha=\alpha^\prime+i\alpha^{\prime\prime}$, and forms a complete set, $ \int \mathrm{d}^2 \alpha\ket{\alpha}\bra{\alpha}/\pi=1$. 
Having said that, tracing out the polaritonic degrees of freedom corresponds to defining an effective Hamiltonian so that 
\begin{equation}\label{eq:Z_decomp_SM}
	Z= \sum_{matter}\bra{matter}\exp(-\beta H_\mathrm{eff})\ket{matter}\;,
\end{equation}
with 
\begin{equation}
	\exp(-\beta H_\mathrm{eff}) = Z_\alpha=\frac{1}{\pi}\int\mathrm{d}^2\alpha
	\bra{\alpha}\exp(-\beta H)\ket{\alpha}\;.\label{eq:Heff_def_single_SM}
\end{equation}
With the help of \autoref{eq:Htot_SM}, we obtain~\cite{coleman_introduction_2015_SM,roman-roche_effective_2022_SM}
\begin{align}
	Z_{\alpha}=\int\dfrac{\mathrm{d}^2\alpha}{\pi}  \exp\left\{-\beta\left[H_{le}+\hbar\omega |\alpha|^2+\sum_{i}\widehat{\boldsymbol{\mu}}_i\cdot\mathbf{E}(\mathbf{r}_i)\alpha+\sum_{i}\widehat{\boldsymbol{\mu}}_i\cdot\mathbf{E}^\ast(\mathbf{r}_i)\alpha^\ast\right]\right\}\;,
\end{align}
or 
\begin{align}
	Z_\alpha=\int \dfrac{\mathrm{d}\alpha^\prime \mathrm{d}\alpha^{\prime\prime}}{{\pi}}  \exp\left[-\beta\left\{H_{le}+\hbar\omega( {\alpha^\prime}^2+{\alpha^{\prime\prime}}^2)+2\sum_{i}\mathrm{Re}\left[ \widehat{\boldsymbol{\mu}}_i\cdot\mathbf{E}(\mathbf{r}_i) \right]\alpha^\prime -2\sum_{i}\mathrm{Im}\left[ \widehat{\boldsymbol{\mu}}_i\cdot\mathbf{E}(\mathbf{r}_i) \right]\alpha^{\prime\prime}\right\}\right]\;.
\end{align}
The integrals are Gaussian and, using their analytical solution, we get
\begin{align}
	Z_{\alpha}& =\dfrac{1}{{\beta\hbar\omega}} \exp\left(\beta H_{le}+ \beta\dfrac{\left\{ \sum_{i}\mathrm{Re}\left[ \widehat{\boldsymbol{\mu}}_i\cdot\mathbf{E}(\mathbf{r}_i) \right]\right\}^2+\left\{\sum_{i}\mathrm{Im}\left[ \widehat{\boldsymbol{\mu}}_i\cdot\mathbf{E}(\mathbf{r}_i) \right]\right\}^2}{\hbar\omega}\right) \nonumber\\
	&=\dfrac{1}{{\beta\hbar\omega}} \exp\left(\beta H_{le}+\beta\sum_{i,j}\widehat{\boldsymbol{\mu}}_i\cdot \dfrac{  \mathrm{Re}\left[\mathbf{E}(\mathbf{r}_i) \otimes\mathbf{E}^\ast(\mathbf{r}_j)\right]}{\hbar\omega}\cdot \widehat{\boldsymbol{\mu}}_j\right)\;,
\end{align}
where  the property $(\sum_i \mathrm{Re}c_i)^2+(\sum_i \mathrm{Im}c_i)^2=\sum_{i,j}(\mathrm{Re}c_i\mathrm{Re}c_j+\mathrm{Im}c_i\mathrm{Im}c_j )=\mathrm{Re}(\sum_{i,j}c_ic_j^\ast)$ was used and $\otimes$ denotes the dyadic product of two vectors.

Using the above expression and \autoref{eq:Heff_def_single_SM}, in the thermodynamic limit~\cite{wang_phase_1973_SM,coleman_introduction_2015_SM,roman-roche_effective_2022_SM} , we obtain the effective Hamiltonian 
\begin{equation}\label{eq:Heff_SM}
	H_\mathrm{eff} =  H_{le}-\sum_{i,j} \widehat{\boldsymbol{\mu}}_i\cdot \Re\left[\dfrac{\mathbf{E}_n(\mathbf{r}_i)\otimes\mathbf{E}^\ast_n(\mathbf{r}_j)}{\hbar\omega}\right]\cdot \widehat{\boldsymbol{\mu}}_j \;.
\end{equation}

For a multi-mode light-matter Hamiltonian
\begin{equation}
	H=H_{le}+\sum_{n}\hbar\omega_n a_n^\dagger a_n +\sum_{i,n}\widehat{\boldsymbol{\mu}}_i\cdot[\mathbf{E}_n(\mathbf{r}_i)a_n+\mathbf{E}_n^\ast(\mathbf{r}_i)a_n^\dagger]\;,
\end{equation}
with the index $n$ denoting the photon modes, the effective Hamiltonian can be obtained in a similar fashion
\begin{equation}\label{eq:Heff_def_multi_SM}
H_\mathrm{eff} =  H_{le}-\sum_{i,j} \widehat{\boldsymbol{\mu}}_i\cdot\left\{ \sum_{n}\Re\left[\dfrac{\mathbf{E}_n(\mathbf{r}_i)\otimes\mathbf{E}^\ast_n(\mathbf{r}_j)}{\hbar\omega_n}\right] \right\}\cdot \widehat{\boldsymbol{\mu}}_j \;.
\end{equation}
The reason is that the photonic modes are not interacting with each other and thus \autoref{eq:Heff_def_single_SM} becomes a product of terms, one for each photon mode.

\section*{Effective Coupling for continuum of EM modes}\label{sec:qed_SM}
Let us now switch from the discrete-mode analysis to a continuum one.
To do so, we employ the macroscopic quantum electrodynamic method, in which the quantized electric field is expressed as 
\begin{equation}
	{\mathbf{E}} (\mathbf{r})= \sum_{p}\int_0^\infty \mathrm{d}\omega\int \mathrm{d}^3r'{\mathbf{G}}_p(\mathbf{r}, \mathbf{r}', \omega){\mathbf{f}}_p( \mathbf{r}', \omega) + \mathrm{H.c.}\label{eq:efield_qed_SM}\;,
\end{equation}
with ${\mathbf{f}}_p(\mathbf{r}', \omega)$ the bosonic annihilation operators of the EM modes, $p$ an index labeling the electric and magnetic contributions and ${\mathbf{G}}_p(\mathbf{r}, \mathbf{r}', \omega)$ functions related to the (classical) dyadic Green's function of Maxwell's equations~\cite{buhmann_dispersion_2012_SM}. 
The electromagnetic interaction of the EM fields with matter in the long-wavelength limit is 
\begin{equation}
	{H}_\mathrm{int}=\sum_{i}\sum_{p}\int_0^\infty \mathrm{d}\omega\int \mathrm{d}^3r'\widehat{\boldsymbol{\mu}}_{i}\cdot[{\mathbf{G}}_p(\mathbf{r}, \mathbf{r}', \omega){\mathbf{f}}_\lambda( \mathbf{r}', \omega)]+\mathrm{H.c.}\;.\label{eq:Hemp_qed_SM}
\end{equation}

The effective matter-matter coupling, $\boldsymbol{\lambda}_{ij}$, can be expressed in terms of the Green's functions by formally discretizing the continuum of cavity modes and expressing it through a collective index $m=\{p,l,\mathbf{r'},\omega\}$ (where $l$ denotes Cartesian components), 
\begin{equation}
	\!\!\!\!\lambda_{ij}^{kk'} = \sum_{m} \dfrac{{E}_m^k(\mathbf{r}_i){E}^{k^\prime \ast}_m(\mathbf{r}_j)}{\hbar\omega_m}= \sum_n \dfrac{1}{\hbar\omega_n}\sum_{p, l}\int\mathrm{d}^3r' G_p^{kl}(\mathbf{r}_i,\mathbf{r}',\omega_n)[G_p^{k'l}(\mathbf{r}_{j},\mathbf{r}',\omega_n)]^{\star\mathrm{T}}=\sum_{n}\dfrac{\omega_n }{\pi\epsilon_0c^2}\Im G^{kk'}(\mathbf{r}_i,\mathbf{r}_{j},\omega_n),
\end{equation}
where $k,k^\prime$ denote Cartesian components. 
In the derivation we used the property $\sum_{p}\int \mathrm{d}^3r' {\mathbf{G}}_p(\mathbf{r}_i,\mathbf{r}',\omega)({\mathbf{G}}_p(\mathbf{r}_{j},\mathbf{r}',\omega))^{\star\mathrm{T}}=\hbar\omega^2\Im{\mathbf{G}}(\mathbf{r}_i,\mathbf{r}_{j},\omega)/(\pi\epsilon_0c^2)$. 
Here $\epsilon_0$ is the vacuum electric permittivity and $c$ is the speed of light in vacuum. 
After taking the continuum limit, we obtain
\begin{equation}
	\boldsymbol{\lambda}_{ij} = \dfrac{1}{\pi\epsilon_0c^2}\int_0^\infty \mathrm{d}\omega \omega \Im{\mathbf{G}}(\mathbf{r}_i,\mathbf{r}_{j},\omega)\label{eq:lambda_gf_SM}\;.
\end{equation}

\section*{Cutoff-dependence}\label{sec:cutoff_SM}

The effective coupling in \autoref{eq:lambda_gf} is formally obtained by an
integral of the frequencies over the whole real axis, which allows the use of
the residue theorem by contour integration. However, it is not obvious that the
upper integration limit should be allowed to go to infinity given that the
long-wavelength (or dipole) approximation was used in the light-matter
interaction, which breaks down when the wavelength becomes comparable to the
extension of the matter components. In this section, we show that the integrals
actually converge for relatively small cutoff frequencies, and thus that setting
the upper integration limit to infinity is a valid approximation. We do this
explicitly for the case of free space, where all calculations are analytical.
The free-space Green's function is given by
\begin{equation}
	\mathbf{G}_0(\mathbf{r},\mathbf{r}',\omega) = \frac{e^{i k \rho}}{4\pi \rho}
	\left[\mathbf{I} - \mathbf{n}_{\rho} \otimes \mathbf{n}_{\rho} +
	\frac{(i k \rho - 1)}{k^2 \rho^2} (\mathbf{I} - 3 \mathbf{n}_{\rho} \otimes \mathbf{n}_{\rho}) \right],
\end{equation}
where $k = \omega/c$, $\rho = |\mathbf{r}-\mathbf{r}'|$, and $\mathbf{n}_{\rho}
= (\mathbf{r}-\mathbf{r}')/\rho$. Inserting this into \autoref{eq:lambda_gf}
immediately yields the kernel of the free-space electrostatic dipole-dipole interaction energy
\begin{equation}\label{eq:free_space_dd_SM}
	\dfrac{1}{2\epsilon_0c^2} \left[\omega^2 \mathbf{G}_0(\mathbf{r}_i,\mathbf{r}_{j},\omega)\right]_{\omega=0} = \frac12 \frac{3 \mathbf{n}_{\rho} \otimes \mathbf{n}_{\rho} - \mathbf{I}}{4\pi \epsilon_0 \rho^3},
\end{equation}
where the factor $1/2$ is due to the fact that each pair of dipoles appears
twice in the sum over $i$ and $j$. We next evaluate the integral in
\autoref{eq:lambda_gf_SM} with an exponential cutoff at wavelength $\Lambda$, i.e.,
frequency $\omega_c = 2\pi c/\Lambda$. To simplify the expressions, we multiply
by the denominator $8\pi\epsilon_0 \rho^3$ of \autoref{eq:free_space_dd_SM},
yielding
\begin{multline}
	8\pi\epsilon_0 \rho^3 \tilde{\boldsymbol{\lambda}}_{ij}(\omega_c)
	= \dfrac{8 \rho^3}{c^2} \int_0^\infty \mathrm{d}\omega \omega \Im{\mathbf{G}_0}(\mathbf{r}_i,\mathbf{r}_{j},\omega) e^{-\omega^2/\omega_c^2} = \\
	4 \pi^{5/2} \rho_\Lambda^3 e^{-\pi^2 \rho_\Lambda^2} \left(\mathbf{I} - \mathbf{n}_{\rho} \otimes \mathbf{n}_{\rho}\right) + \left(1 - \operatorname{erfc}(\pi \rho_\Lambda) - 2\sqrt{\pi} \rho_\Lambda e^{-\pi^2 \rho_\Lambda^2}\right) \left(3\mathbf{n}_{\rho} \otimes \mathbf{n}_{\rho} - \mathbf{I}\right),
\end{multline}
where we have introduced $\rho_\Lambda = \rho/\Lambda$ for simplicity. In the limit $\Lambda\to 0$ (i.e., $\omega_c \to \infty$), this clearly recovers \autoref{eq:free_space_dd_SM}. Interestingly, this convergence is very fast, as all terms apart from the surviving constant one are suppressed with $e^{-\pi^2 x^2}$ (which is also the asymptotic behavior of $\operatorname{erfc}$). For $\rho_\Lambda = 1$, the relative error is on the level of $0.4\%$, and for $\rho_\Lambda = 2$, it is already below $10^{-14}$. This implies that for a given distance $\rho$ between points, a cutoff wavelength similar to that distance is sufficient to obtain a fully converged result. Since the spatial extent of the matter components can at most be of the order of their distance (and is typically significantly smaller), this implies that the dipole approximation is expected to work well. The formal extension of the frequency integral to an upper limit of infinity without going beyond the dipole approximation is thus well-justified. We note that while we have here studied the case of free space, the same arguments are expected to apply in general environments.

\end{document}